\documentclass[11pt,a4paper]{article}
\usepackage{jcappub}

\usepackage[latin1]{inputenc}
\usepackage{graphicx}
\usepackage{epsfig}
\usepackage{color}
\usepackage{comment}
\usepackage{slashed}

\title{Robust Limits on Photon Mass from Statistical Samples of Extragalactic Radio Pulsars}

\author[a,b]{Jun-Jie Wei,}
\author[a,c]{Xue-Feng Wu}

\affiliation[a]{Purple Mountain Observatory, Chinese Academy of Sciences, Nanjing 210008, China}
\affiliation[b]{Guangxi Key Laboratory for Relativistic Astrophysics, Guangxi University, Nanning 530004, China}
\affiliation[c]{School of Astronomy and Space Sciences, University of Science and Technology of China, Hefei 230026, China}

\emailAdd{jjwei@pmo.ac.cn; xfwu@pmo.ac.cn}

\abstract{The photon zero-mass hypothesis has been investigated for a long time using the frequency-dependent time delays
of radio emissions from astrophysical sources. However, the search for a rest mass of the photon has been hindered by the similarity between
the frequency-dependent dispersions due to the plasma and nonzero photon mass effects.
Considering the contributions to the observed dispersion measure from both the plasma and nonzero photon mass effects,
and assuming the dispersion induced by the plasma effect is an unknown constant,
we obtain a robust limit on the photon mass by directly fitting a combination of the dispersion measures of radio sources.
Using the observed dispersion measures from two statistical samples of extragalactic pulsars,
here we show that at the 68\% confidence level, the constraints on the photon mass can be as low as
$m_{\gamma}\leq1.51\times10^{-48}~\rm kg\simeq8.47\times10^{-13}~{\rm eV}/c^{2}$ for the sample of 22 radio pulsars
in the Large Magellanic Cloud and $m_{\gamma}\leq1.58\times10^{-48}~\rm kg\simeq8.86\times10^{-13} {\rm eV}/c^{2}$
for the other sample of 5 radio pulsars in the Small Magellanic Cloud, which are comparable with
that obtained by a single extragalactic pulsar. Furthermore, the statistical approach presented here can also be used
when more fast radio bursts with known redshifts are detected in the future.}


\arxivnumber{1803.07298}

\begin{document}
\maketitle

 \flushbottom


\section{Introduction}

A basic assumption of Maxwell's electromagnetism as well as Einstein's special relativity
is the constant speed $c$, in a vacuum, of all electromagnetic radiation. That is, the photon
is expected to be massless. One of the most enduring efforts on testing the correctness
of this assumption has therefore been the search for a rest mass of the photon.
However, it is impossible to prove experimentally that the photon rest mass
is strictly zero, the best experimentally strategy can hope to do is to set ever tighter upper limits on it
and push the verification of the photon zero-mass hypothesis as far as possible.
Based on the uncertainty principle, when using the age of the universe ($\sim10^{10}$ yr),
the ultimate upper limit on the photon mass is estimated to be
$m_{\gamma}\approx\hbar/Tc^2\approx10^{-69}~\rm{kg}$ \cite{1971RvMP...43..277G,2005RPPh...68...77T}.
Although such an infinitesimal upper limit would be extremely
difficult to place, there are several possible visible effects associated with a nonzero photon rest mass.
These effects have been employed to set upper bounds on the photon mass via different
terrestrial and astronomical approaches \cite{1971RvMP...43..277G,1973PhRvD...8.2349L,2005RPPh...68...77T,2006AcPPB..37..565O,2010RvMP...82..939G,2011EPJD...61..531S}.

To date, the methods for constraining the photon mass include measurement of the frequency dependence in the velocity of light \cite{1964Natur.202..377L,1969Natur.222..157W,1999PhRvL..82.4964S,2016ApJ...822L..15W,2016PhLB..757..548B,2017PhLB..768..326B,2016JHEAp..11...20Z,2017RAA....17...13W,2017PhRvD..95l3010S},
tests of Coulomb's inverse square law \cite{1971PhRvL..26..721W}, tests of Amp$\rm \grave{e}$re's law \cite{1992PhRvL..68.3383C},
gravitational deflection of electromagnetic radiation \cite{1973PhRvD...8.2349L,2004PhRvD..69j7501A},
Jupiter's magnetic field \cite{1975PhRvL..35.1402D}, mechanical stability of the magnetized gas in galaxies \cite{1976UsFiN.119..551C},
torsion Cavendish balance \cite{1998PhRvL..80.1826L,2003PhRvL..91n9101G,2003PhRvL..90h1801L,2003PhRvL..91n9102L},
magnetohydrodynamic phenomena of the solar wind \cite{1997PPCF...39...73R,2007PPCF...49..429R,2016APh....82...49R},
black hole bombs \cite{2012PhRvL.109m1102P}, pulsar spindown \cite{2017ApJ...842...23Y}, and so on.
Among these methods, the most direct and model-independent one is to measure the frequency dependence
in the speed of light. In this paper, we will revisit the photon mass limits from
the velocity dispersion of electromagnetic waves of astronomical sources.

If the photon mass is nonzero ($m_\gamma\neq0$), the energy of the photon can be written as
\begin{equation}\label{eq1}
E=h\nu=\sqrt{p^2c^2+m_\gamma^2c^4}\;.
\end{equation}
The dispersion relation between the group velocity of photon $\upsilon$
and frequency $\nu$ is
\begin{equation}\label{eq2}
\upsilon=\frac{\partial{E}}{\partial{p}}=c\sqrt{1-\frac{m_\gamma^2c^4}{E^2}}\approx c\left(1-\frac{1}{2}\frac{m_\gamma^2c^4}{h^2\nu^2}\right)\;,
\end{equation}
where the last derivation holds when $m_\gamma\ll h\nu/c^{2}\simeq7\times10^{-42}\left(\frac{\nu}{\rm GHz}\right)\;{\rm kg}$.
It is obvious from Equation~(\ref{eq2}) that lower frequency photons would propagate slower than
higher frequency ones in a vacuum. The arrival-time differences of photons with different energies
originating from the same source can therefore be used to constrain the photon mass. Moreover,
it is easy to understand that measurements of short time structures at lower frequencies from
distant astronomical sources are especially powerful for constraining the photon mass.
The current best limits on the photon mass through the dispersion of light have been made using
the radio emissions from fast radio bursts (FRBs; $m_{\gamma}\leq3.9\times10^{-50}~\rm kg$)
\cite{2016ApJ...822L..15W,2016PhLB..757..548B,2017PhLB..768..326B} and pulsars in the Large and
Small Magellanic Clouds (LMC and SMC respectively; $m_{\gamma}\leq2.3\times10^{-48}~\rm kg$)
\cite{2017RAA....17...13W}. Recently, Ref.~\cite{2017PhRvD..95l3010S} extended previous studies to FRBs
where the redshift is not available, and they constructed a Bayesian formula to measure
the photon mass with a catalog of FRBs.

However, it is well known that radio signals propagating through a plasma would arrive with a frequency-dependent
dispersion in time of the $\nu^{-2}$ behavior. This is the frequency dependence expected from the plasma
effect, but a similar dispersion $\propto m_\gamma^2/\nu^2$ (see Equation~\ref{eq2}) could also arise from
a nonzero photon mass. In other words, the dispersion method used for testing the photon mass is hindered
by the similar frequency-dependent dispersions from the plasma and photon mass.
In order to identify an effect as radical as a massive photon, statistical
and possible systematic uncertainties must be minimized. One cannot rely on a single source, for which
it would be impossible to distinguish the dispersions induced by the plasma effect and the nonzero photon
mass effect. For this reason, we develop a statistical method through which an average dispersion
measure caused by the the plasma effect can be extracted and a combined constraint on the photon rest mass
can be obtained as well, by fitting a combination of the dispersion measures of astronomical sources.

In this work, we gather the observed dispersion measures from two samples of radio pulsars in the LMC and SMC
to constrain the photon mass. The paper is organized as follows. In Section~\ref{sec:method},
we give an overview of the theoretical analysis framework. The resulting constraints on the photon mass are
presented in Section~\ref{sec:results}. Our conclusions are summarized in Section~\ref{sec:summary}.

\section{Theoretical Framework}
\label{sec:method}

As described above, the pulse arrival time delay at a given frequency $\nu$ follows the $\nu^{-2}$ law.
In this work, we suppose that the frequency-dependent delay can be attributed to two causes: (i) the plasma effect via
the dispersion process from the line-of-sight free electron content, and (ii) the nonzero photon mass (if it exits).

\subsection{Dispersion from the plasma effect}
Due to the dispersive nature of plasma, lower frequency radio waves pass through the line-of-sight free electrons
slower than higher frequency ones \cite{2017AdSpR..59..736B}. The arrival time delay ($\Delta t_{\rm DM}$) between
two pulses with different frequencies ($\nu_{l}<\nu_{h}$), which induced by the plasma effect, is expressed as
\begin{equation}\label{tDM}
\begin{aligned}
  \Delta{t_{\rm DM}}&=\int \frac{{\rm d}l}{c} \frac{\nu^{2}_{p}}{2}\left(\nu_l^{-2}-\nu_h^{-2}\right)\\
  &=\frac{e^{2}}{8\pi^{2} m_{e}\epsilon_{0}c}\left(\nu_l^{-2}-\nu_h^{-2}\right){\rm DM_{astro}}\;,
\end{aligned}
\end{equation}
where $\nu_{p}=(n_{e}e^{2}/4\pi^{2} m_{e}\epsilon_{0})^{1/2}$ is the plasma frequency, $n_{e}$ is
the number density of electrons, $e$ and $m_{e}$ are the charge and mass of the electron, and
$\epsilon_{0}$ is the permittivity of vacuum. The dispersion measure ($\rm DM_{astro}$) is defined as
the integral of the electron number density along the propagation path from the source to the observer,
${\rm DM_{astro}}\equiv\int n_{e}{\rm d}l$.

For a Magellanic Cloud pulsar, the $\rm DM_{astro}$ has contributions from the electron density
in our Galaxy ($\rm DM_{Gal}$), the Magellanic Cloud ($\rm DM_{MC}$), and the intergalactic medium
between the two galaxies ($\rm DM_{IGM}$). In fact, since each galaxy has an extended circum-galaxy medium,
there might not be a clear separation between the two. In any case, all these complications do not enter the
problem we are treating since only $\rm DM_{astro}$ is relevant.
We will present in Section~\ref{subsec:likelihood} how $\rm DM_{astro}$ is treated in a maximum
likelihood estimation.

\subsection{Dispersion from a nonzero photon mass}

With Equation~(\ref{eq2}), it is evident that two photons with different frequencies originating from
the same source would arrive on Earth at different times. The arrival time difference due to a nonzero
photon mass is given by
\begin{equation}\label{tmr}
  \Delta{t_{m_{\gamma}}}=\frac{Lm_\gamma^2c^3}{2h^2}\left(\nu_l^{-2}-\nu_h^{-2}\right)\;,
\end{equation}
where $L$ is the distance of the source.

\subsection{Analysis method}
\label{subsec:likelihood}

In our analysis, the observed time delay ($\Delta{t_{\rm obs}}$) between correlated photons should consist of two terms
\begin{equation}\label{tobs}
\Delta{t_{\rm obs}}=\Delta{t_{\rm DM}}+\Delta{t_{m_{\gamma}}}\;.
\end{equation}
These two terms have the same frequency dependence, which are consistent with the observational fact that
the observed time delay $\Delta{t_{\rm obs}}\propto\nu^{-2}$.
Note that the observed dispersion measure ($\rm DM_{obs}$) is directly derived from the fitting of the $\nu^{-2}$
behavior of the observed time delay. That is, both the line-of-sight free electron content and a massive photon
determine the same $\rm DM_{obs}$, i.e.,
\begin{equation}\label{DMobs}
{\rm DM_{obs}}={\rm DM_{astro}}+{\rm DM_{\gamma}}\;,
\end{equation}
where $\rm DM_{asro}$ is the dispersion due to the plasma effect and $\rm DM_{\gamma}$ denotes the ``effective dispersion
measure'' induced by a nonzero photon mass, defined by \cite{2017PhRvD..95l3010S}
\begin{equation}\label{DMr}
{\rm DM_{\gamma}}\equiv\frac{4\pi^2m_e\epsilon_{0}c^4}{h^2e^2}Lm_\gamma^2\;.
\end{equation}

Statistically, a combined constraint on the photon mass $m_{\gamma}$ can be obtained by combining
results of several measurements of the dispersion measure into an overall result. That is,
for a set of extragalactic radio pulsars, the photon mass $m_{\gamma}$ can be constrained
by maximizing the joint likelihood function
\begin{equation}\label{likelihood}
\begin{aligned}
\mathcal{L} = &\prod_{i}
\frac{1}{\sqrt{2\pi\left(\sigma_{\rm DM_{obs,\emph{i}}}^{2}+\sigma_{\rm DM_{astro,\emph{i}}}^{2}\right)}}\\
&\times\exp\left[-\,\frac{\left(\rm DM_{obs,\emph{i}}-DM_{astro,\emph{i}}-DM_{\gamma}\right)^{2}}
{2\left(\sigma_{\rm DM_{obs,\emph{i}}}^{2}+\sigma_{\rm DM_{astro,\emph{i}}}^{2}\right)}\right]\;,
      \end{aligned}
\end{equation}
where $i$ is the corresponding serial number of each pulsar, $\sigma_{\rm DM_{obs}}$ and $\sigma_{\rm DM_{astro}}$
are respectively the uncertainty of $\rm DM_{obs}$ and the uncertainty of $\rm DM_{astro}$.

Generally, given a model for the distribution of free electrons in our Galaxy, the Magellanic Cloud
and the intergalactic medium between the two galaxies, we can estimate the contributions of $\rm DM_{Gal}$,
$\rm DM_{MC}$ and $\rm DM_{IGM}$ to the $\rm DM_{astro}$.
However, the knowledge about the electron distributions is poor, and the current electron-density model
relies on many particular theoretical parameters \cite{2017ApJ...835...29Y}. It is obviously that
the estimation of $\rm DM_{astro}$ is model-dependent and its uncertainty $\sigma_{\rm DM_{astro}}$
is large. For simplicity, we will adopt an average dispersion measure $\langle \rm DM_{astro}\rangle$
and use it uniformly for every radio pulsar in the same Magellanic Cloud. To account for possible
electron-density model inaccuracy, we will adopt the additional free parameter $\eta$ to relate
the uncertainty $\sigma_{\rm DM_{astro}}$ to the observed dispersion measure $\rm DM_{obs}$,
according to $\sigma_{\rm DM_{astro}}\equiv\eta\rm DM_{astro}\simeq\eta\rm DM_{obs}$
(where $\rm DM_{\gamma} \ll DM_{astro}$). The Particle Data Group suggests the currently adopted
upper limit of photon mass is $m_{\gamma}\leq1.5\times10^{-54}$ kg \cite{2014ChPhC..38i0001O}.
With such a limit, $\rm DM_{\gamma}$ can be estimated as $\sim 10^{-11}$ pc $\rm cm^{-3}$ at a distance
of 50 kpc (see Equation~\ref{DMr}), which is obviously far less than the value of $\rm DM_{astro}$.
Thus, it is reasonable to assume $\rm DM_{\gamma} \ll DM_{astro}$.
With these treatments above, the likelihood function becomes
\begin{equation}\label{likelihood2}
\begin{aligned}
\mathcal{L} = &\prod_{i}
\frac{1}{\sqrt{2\pi\left(\sigma_{\rm DM_{obs,\emph{i}}}^{2}+\eta^2\rm DM_{obs,\emph{i}}^{2}\right)}}\\
&\times\exp\left[-\,\frac{\left(\rm DM_{obs,\emph{i}}-\langle DM_{astro}\rangle-DM_{\gamma}\right)^{2}}
{2\left(\sigma_{\rm DM_{obs,\emph{i}}}^{2}+\eta^2\rm DM_{obs,\emph{i}}^{2}\right)}\right]\;.
      \end{aligned}
\end{equation}
We will use the likelihood in Equation~(\ref{likelihood2}) to simultaneously constrain
the photon mass $m_{\gamma}$ and the parameters $\langle \rm DM_{astro}\rangle$ and $\eta$.

\begin{table}
\caption{Radio pulsars in the Magellanic Clouds$^{\rm a}$}
\footnotesize
\centering
\begin{tabular}{lccccc}
\hline
\hline
 J2000 &  Cloud  &   R.A.         &   Decl.          &   $\rm DM_{obs}$          & Refs. \\
 Name  &         &  ($^{\circ}$)  &   ($^{\circ}$)   & (pc cm$^{-3}$) & \\
\hline
J0045--7042	&	SMC	&	11.357	&	-70.702	&$	70	\pm	3	$&	\cite{2006ApJ...649..235M}\\
J0045--7319	&	SMC	&	11.397	&	-73.317	&$	105.4	\pm	7	$&	\cite{1991MNRAS.249..654M,1994ApJ...423L..43K}\\
J0111--7131	&	SMC	&	17.870	&	-71.530	&$	76	\pm	3	$&	\cite{2006ApJ...649..235M}\\
J0113--7220	&	SMC	&	18.296	&	-72.342	&$	125.49	\pm	3	$&	\cite{2001ApJ...553..367C}\\
J0131--7310	&	SMC	&	22.869	&	-73.169	&$	205.2	\pm	7	$&	\cite{2006ApJ...649..235M}\\
\hline
J0449--7031	&	LMC	&	72.274	&	-70.525	&$	65.83	\pm	7	$&	\cite{2006ApJ...649..235M}\\
J0451--67	&	LMC	&	72.958	&	-67.300	&$	45	\pm	1	$&	\cite{2006ApJ...649..235M}\\
J0455--6951	&	LMC	&	73.948	&	-69.860	&$	94.89	\pm	14	$&	\cite{1991MNRAS.249..654M,2001ApJ...553..367C}\\
J0456--69	&	LMC	&	74.125	&	-69.167	&$	103	\pm	1	$&	\cite{2013MNRAS.433..138R}\\
J0456--7031	&	LMC	&	74.010	&	-70.519	&$	100.3	\pm	3	$&	\cite{2006ApJ...649..235M}\\
J0457--69	&	LMC	&	74.258	&	-69.767	&$	91	\pm	1	$&	\cite{2013MNRAS.433..138R}\\
J0458--67	&	LMC	&	74.746	&	-67.717	&$	97	\pm	2	$&	\cite{2013MNRAS.433..138R}\\
J0502--6617	&	LMC	&	75.711	&	-66.300	&$	68.9	\pm	3	$&	\cite{1991MNRAS.249..654M,2001ApJ...553..367C}\\
J0519--6932	&	LMC	&	79.945	&	-69.540	&$	119.4	\pm	5	$&	\cite{2006ApJ...649..235M}\\
J0521--68	&	LMC	&	80.433	&	-68.583	&$	136	\pm	4	$&	\cite{2013MNRAS.433..138R}\\
J0522--6847	&	LMC	&	80.596	&	-68.784	&$	126.45	\pm	7	$&	\cite{2006ApJ...649..235M}\\
J0529--6652	&	LMC	&	82.462	&	-66.877	&$	103.2	\pm	3	$&	\cite{1983Natur.303..307M,2001ApJ...553..367C}\\
J0532--6639	&	LMC	&	83.248	&	-66.660	&$	69.3	\pm	18	$&	\cite{2006ApJ...649..235M}\\
J0532--69	&	LMC	&	83.017	&	-69.767	&$	124	\pm	1	$&	\cite{2013MNRAS.433..138R}\\
J0534--6703	&	LMC	&	83.651	&	-67.064	&$	94.7	\pm	12	$&	\cite{2006ApJ...649..235M}\\
J0535--66	&	LMC	&	83.917	&	-66.867	&$	75	\pm	1	$&	\cite{2013MNRAS.433..138R}\\
J0535--6935	&	LMC	&	83.750	&	-69.583	&$	93.7	\pm	4	$&	\cite{2001ApJ...553..367C,2006ApJ...649..235M}\\
J0537--69	&	LMC	&	84.429	&	-69.350	&$	273	\pm	1	$&	\cite{2013MNRAS.433..138R}\\
J0540--6919	&	LMC	&	85.047	&	-69.332	&$	146.5	\pm	2	$&	\cite{1984ApJ...287L..19S,2003ApJ...590L..95J}\\
J0542--68	&	LMC	&	85.646	&	-68.267	&$	114	\pm	5	$&	\cite{2013MNRAS.433..138R}\\
J0543--6851	&	LMC	&	85.970	&	-68.857	&$	131	\pm	4	$&	\cite{2006ApJ...649..235M}\\
J0555--7056	&	LMC	&	88.758	&	-70.946	&$	73.4	\pm	16	$&	\cite{2006ApJ...649..235M}\\
\hline
\end{tabular}
\label{table1}
\\
$^{\rm a}$http://www.atnf.csiro.au/research/pulsar/psrcat \cite{2005AJ....129.1993M}
\end{table}

\section{Constraints on Photon Mass}
\label{sec:results}

Up to now, the 29 known extragalactic pulsars are all in the Magellanic Clouds.
Only 27 of these have known dispersion measures, with one pulsar in each of
the LMC and SMC being detected at high energies and having no radio counterpart.
Of these 27 radio pulsars, 22 in the LMC and 5 in the SMC. Their observed
dispersion measures $\rm DM_{obs}$ are listed in Table~\ref{table1}
along with their location information (including the right ascension coordinate
and the declination coordinate). The distances of the LMC and SMC are 49.7 and 59.7 kpc,
respectively. Since these extragalactic pulsars appear to be usually located in
the more central regions of each Magellanic Cloud \cite{2006ApJ...649..235M},
we adopt the distance of the corresponding Cloud as the distance of each pulsar which lies in it.

For each sample of radio pulsars in the LMC and SMC, we use the Python Markov Chain Monte Carlo (MCMC)
module, EMCEE \cite{2013PASP..125..306F}, to explore the posterior distributions of parameters
($m_{\gamma}$, $\langle \rm DM_{astro}\rangle$, and $\eta$). We choose uniform priors on the parameters,
such that $10^{-69}<m_{\gamma}/{\rm kg}<10^{-42}$,\footnote{The lower end is adopted because it
corresponds to the ultimate upper limit estimated by the uncertainty principle, while the upper end
is adopted because beyond which the approximation in Equation~(\ref{eq2}) breaks down.}
$0.0<\langle \rm DM_{astro}\rangle/[{\rm pc\;cm^{-3}}]<300$, and $0.0<\eta<1.0$.

\begin{figure}
   \centering
  \includegraphics[width=85mm]{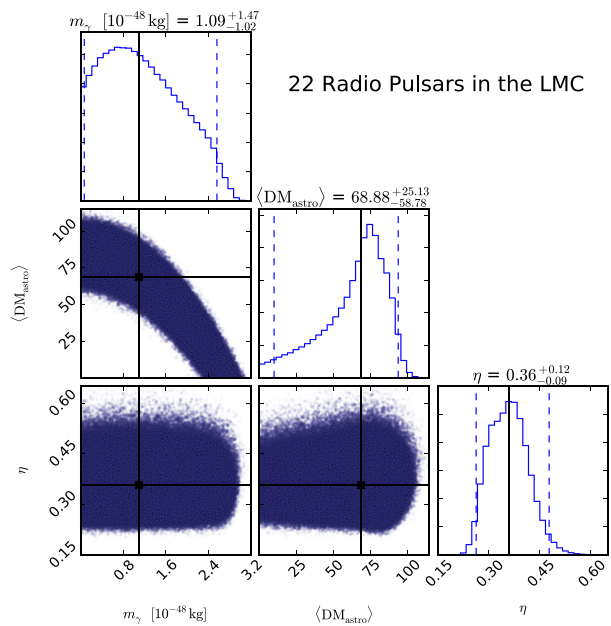}
  \vskip-0.2in
\caption{1D marginalized probability distributions and 2D regions corresponding to
the photon mass $m_{\gamma}$ and the parameters $\langle \rm DM_{astro}\rangle$ and $\eta$,
using the combination of 22 radio pulsars in the LMC. The vertical solid lines represent the central values,
and the vertical dashed lines enclose the 95\% credible region.}
\label{fig1}
\end{figure}

\begin{figure}
   \centering
  \includegraphics[width=85mm]{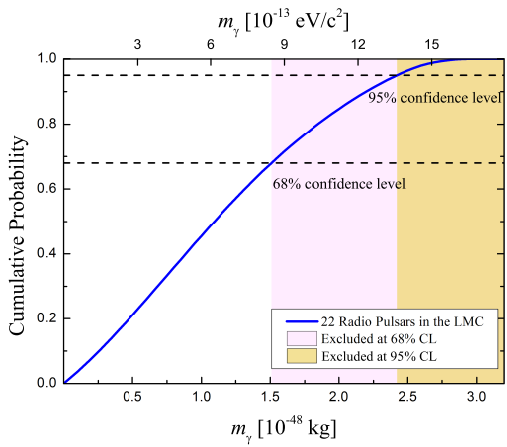}
  \vskip-0.2in
\caption{The cumulative posterior probability distribution on $m_{\gamma}$ from
the sample of radio pulsars in the LMC. The excluded values of $m_{\gamma}$
at 68\% and 95\% CLs are displayed with shaded areas.}
\label{fig2}
\end{figure}

\subsection{Limit from a sample of radio pulsars in the LMC}

In Figure~\ref{fig1}, we show the marginalized posterior probability densities of $m_{\gamma}$,
$\langle \rm DM_{astro}\rangle$, and $\eta$ for the combination of 22 radio pulsars in the LMC.
One can see from this plot that at the 95\% confidence level, the central values and the
corresponding uncertainties of the parameters are $m_{\gamma}=(1.09^{+1.47}_{-1.02})\times10^{-48}$ kg,
$\langle \rm DM_{astro}\rangle=68.88^{+25.13}_{-58.78}$ pc cm$^{-3}$, and $\eta=0.36^{+0.12}_{-0.09}$, respectively.
Note that the derived parameters and their error bars are based on their respective marginalized distributions: the central values
correspond to the 50th percentile in the marginalized distributions, while their error bars correspond to the
$50^{th}-2.5^{th}$ and $97.5^{th}-50^{th}$ percentiles of the marginalized distributions.
Figure~\ref{fig2} shows the marginalized accumulative posterior probability distribution on $m_{\gamma}$.
The 68\% and 95\% confidence-level upper limits on $m_{\gamma}$ are
\begin{equation}
m_{\gamma}\leq1.51\times10^{-48} {\rm kg}\simeq8.47\times10^{-13} {\rm eV}/c^{2}
\end{equation}
and
\begin{equation}
m_{\gamma}\leq2.42\times10^{-48} {\rm kg}\simeq1.36\times10^{-12} {\rm eV}/c^{2}\;,
\end{equation}
respectively. These 68\% and 95\% confidence-level limits are comparable with that
obtained by a single LMC pulsar \cite{2017RAA....17...13W}.

\begin{figure}
   \centering
  \includegraphics[width=85mm]{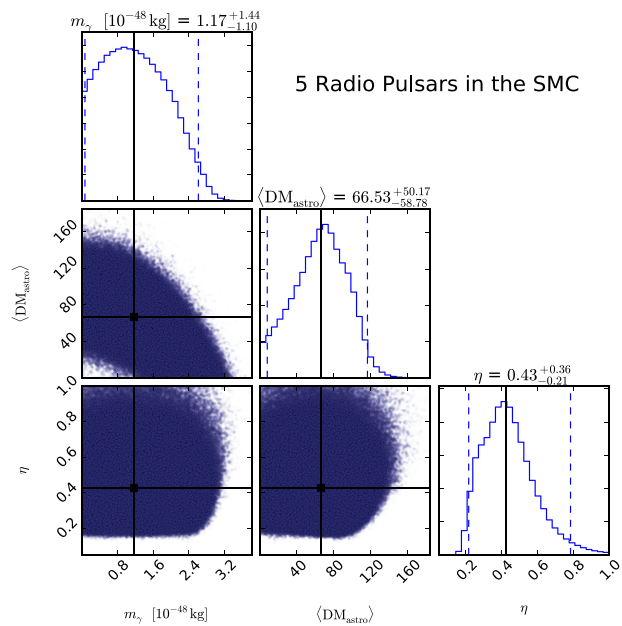}
  \vskip-0.2in
\caption{Same as Figure~\ref{fig1}, except now using the combination of 5 radio pulsars in the SMC.}
\label{fig3}
\end{figure}

\begin{figure}
   \centering
  \includegraphics[width=85mm]{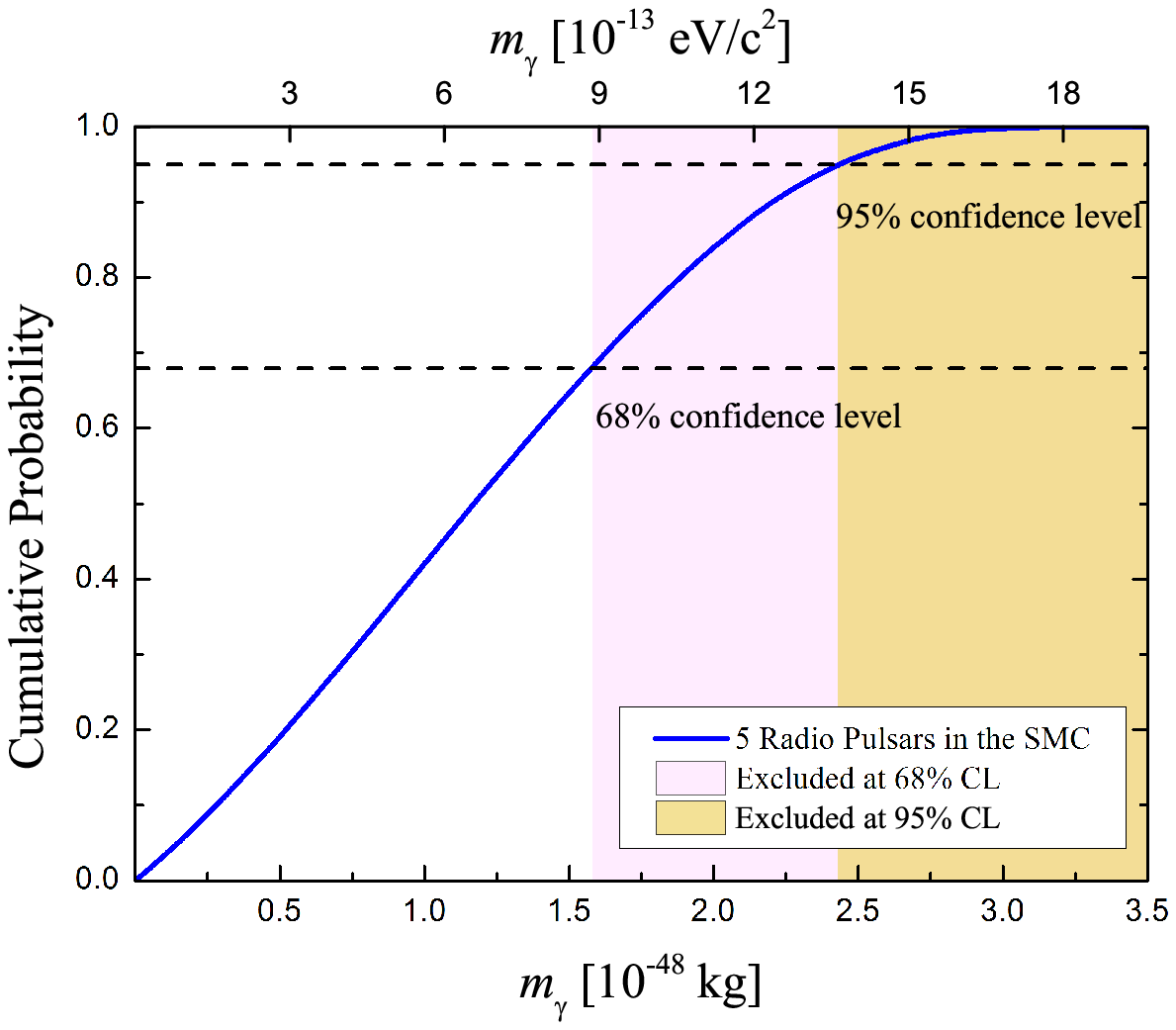}
  \vskip-0.2in
\caption{Same as Figure~\ref{fig2}, but now for the sample of radio pulsars in the SMC.}
\label{fig4}
\end{figure}

\subsection{Limit from a sample of radio pulsars in the SMC}

The marginalized posterior probability distributions for the combination of 5 radio pulsars in the SMC
are presented in Figure~\ref{fig3}. Similarly, we find that the 95\% confidence level constraints on
the parameters are $m_{\gamma}=(1.17^{+1.44}_{-1.10})\times10^{-48}$ kg,
$\langle \rm DM_{astro}\rangle=66.53^{+50.17}_{-58.78}$ pc cm$^{-3}$, and $\eta=0.43^{+0.36}_{-0.21}$, respectively.
 We also display the accumulative posterior
probability distribution of $m_{\gamma}$ in Figure~\ref{fig4}, which implies
\begin{equation}
m_{\gamma}\leq1.58\times10^{-48} {\rm kg}\simeq8.86\times10^{-13} {\rm eV}/c^{2}
\end{equation}
and
\begin{equation}
m_{\gamma}\leq2.43\times10^{-48} {\rm kg}\simeq1.36\times10^{-12} {\rm eV}/c^{2}\;,
\end{equation}
at the 68\% and 95\% confidence levels, respectively.
This 95\% confidence-level limit is also as good as the previous result that
only used a single SMC pulsar \cite{2017RAA....17...13W}.

\section{Summary and Discussion}
\label{sec:summary}

The frequency-dependent time delays of radio emissions from astrophysical sources have been used to constrain the rest mass of the photon
with high accuracy. However, the plasma and nonzero photon mass effects on photon propagation would cause
the similar frequency-dependent dispersions. The key issue in the idea of searching for frequency-dependent delays,
therefore, is distinguishing the dispersions induced by the plasma effect and the nonzero photon mass effect.
Here we develop a statistical method based on the global fitting of
observed dispersion measures $\rm DM_{obs}$ of radio sources to constrain the photon mass, and an unknown constant
is assumed to be the average dispersion measure $\langle \rm DM_{astro}\rangle$ arised from the plasma effect
for every radio source. This is the first approach which can give a combined constraint on the photon mass.

Using the observed dispersion measures from two samples of extragalactic pulsars, we place robust limits
on the photon mass at the 68\% (95\%) confidence level, i.e., $m_{\gamma}\leq1.51\times10^{-48}~\rm kg$ ($m_{\gamma}\leq2.42\times10^{-48}~\rm kg$)
for the sample of 22 radio pulsars in the LMC and $m_{\gamma}\leq1.58\times10^{-48}~\rm kg$ ($m_{\gamma}\leq2.43\times10^{-48}~\rm kg$)
for the other sample of 5 radio pulsars in the SMC. Compared with previous limits from a single pulsar in each of
the LMC and SMC \cite{2017RAA....17...13W}, our 95\% confidence-level constraints from statistical samples
are equally good.

Although our limits on the photon mass are two orders of magnitude worse than the current best limit from
a single FRB \cite{2016ApJ...822L..15W,2016PhLB..757..548B,2017PhLB..768..326B}, there is merit to the results.
First, thanks to our improved statistical technique and the adoption of a more complete data set, our constraints
are much more statistically robust than previous results. Second, the analysis of the dispersion measure $\rm DM_{astro}$
from the plasma effect performed here is important for studying the frequency dependence of the speed of light
to constrain the photon mass, since it impacts the reliability of the resulting constraints on $m_{\gamma}$.
Compared with previous works \cite{2016ApJ...822L..15W,2016PhLB..757..548B,2017PhLB..768..326B,2017PhRvD..95l3010S},
which estimated $\rm DM_{astro}$ based on the given model for the distribution of free electrons in the Galaxy,
the Magellanic Clouds, and the intergalactic medium, our present analysis is independent of the electron-density model.
Furthermore, although extragalactic pulsars are discussed in this work, the approach presented here can also be used
for other radio sources, such as future FRBs with more known redshifts.

\acknowledgments
We are grateful to the anonymous referee for useful comments and suggestions.
This work is partially supported by the National Basic Research Program (``973'' Program)
of China (Grant No. 2014CB845800), the National Natural Science Foundation of China
(Grant Nos. 11603076, 11673068, and 11725314), the Youth Innovation Promotion
Association (2011231 and 2017366), the Key Research Program of Frontier Sciences (Grant No. QYZDB-SSW-SYS005),
the Strategic Priority Research Program ``Multi-waveband gravitational wave Universe''
(Grant No. XDB23000000) of the Chinese Academy of Sciences, and the Natural Science Foundation
of Jiangsu Province (Grant No. BK20161096).


\providecommand{\href}[2]{#2}\begingroup\raggedright\endgroup

\end{document}